\documentclass[a4paper,french]{rnti} 

\usepackage[T1]{fontenc}
\usepackage[utf8]{inputenc}
\usepackage{multirow}

\usepackage{graphicx}
\usepackage{subcaption}
\usepackage{float}

\titrecourt{Agro-STAY pour collecter et analyser les données YouTube}

\nomcourt{L. Maxim et al.}

\titre{Agro-STAY : Collecte de données et analyse des informations en agriculture alternative issues de YouTube}

\auteur{Laura Maxim\affil{1},
        Julien Rabatel\affil{2}, 
        Jean-Marc Douguet\affil{3}, 
        Natalia Grabar\affil{4}, \\
        Roberto Interdonato\affil{2}\affilsep\affil{6},
        Sébastien Loustau\affil{5},
        Mathieu Roche\affil{2}\affilsep\affil{6},
        Maguelonne Teisseire\affil{6}
        }

\affiliation{
    \affil{1} LISIS, CNRS, INRAE, Université Gustave Eiffel,
ESIEE Paris\\
          laura.maxim@cnrs.fr,\\
    \affil{2} CIRAD, UMR TETIS, F-34398 Montpellier, France\\
    \affil{3} ESE, Univ. Paris-Saclay, CNRS, INRAE, AgroParisTech, IRD, Gif-sur-Yvette, France\\
    \affil{4} CNRS, Univ. Lille, UMR 8163 - STL - Savoirs Textes Langage, F-59000 Lille, France\\
    \affil{5} Université de Pau et des Pays de l'Adour, France\\
    \affil{6} TETIS, Université de Montpellier, AgroParisTech, CIRAD, INRAE, Montpellier, France
 }

\resume{%
Face aux crises actuelles (climatiques, sociales, économiques), l’autosuffisance - ensemble de pratiques combinant sobriété énergétique, autoproduction alimentaire et énergétique et autoconstruction - suscite un intérêt croissant. 
Le projet CNRS STAY (Savoirs Techniques pour l’Auto-suffisance, sur YouTube) s’inscrit dans ce domaine en analysant les savoirs techniques diffusés sur YouTube. Nous présentons Agro-STAY\footnote{\texttt{http://agro-stay.com:8033/}}, une plateforme dédiée à la collecte, au traitement et à la visualisation de données issues de vidéos YouTube et de leurs commentaires. En mobilisant des techniques de traitement automatique des langues (TAL) et des modèles de langues, ce travail permet une analyse fine des pratiques agricoles alternatives décrites en ligne. 
}

\summary{%
For facing the current crises (climatic, social, economic), the self-sufficiency - a set of practices that combine energy sobriety, self-production of food and energy, and self-construction - arouses an increasing interest. 
The project CNRS STAY (Savoirs Techniques pour l’Auto-suffisance, sur YouTube) addresses this topic through an analysis of techniques publicised on YouTube. We present Agro-STAY, a platform dedicated to the collection, processing and visualization of data issued from YouTube videos and their comments. We exploit techniques from Natural Language Processing (NLP) and language models, which enable a fine-grained analysis of alternative agricultural practice described online. 
}

\begin{document}

\section{Introduction}

Face aux crises récentes, annoncées ou redoutées (changement climatique, guerres, pandémie, accès plus difficile et coûteux aux ressources non renouvelables, perte de biodiversité, effondrements financiers mondiaux...), des comportements adaptatifs se développent chez les individus, visant l’autoproduction alimentaire, l’accès à l’eau potable, l’autoproduction de ressources énergétiques (bois, énergie solaire, eau chaude...) ou encore l'auto-construction des habitations. En associant la sobriété énergétique et la limitation significative des achats de biens de consommation, l'autoproduction s'accompagne des changements dans les modes de consommation dans une approche globale de renforcement de leur résilience.

La littérature sociologique montre que l'autosuffisance peut se situer en référence à des échelles de temps différentes mais cohérentes. A court terme, il s’agirait d'une réponse à une remise en cause profonde d'un style de vie organisée autour d'un travail salarié, qui peut être vécu comme étant dénué de sens, dans des environnements managériaux impactant l’état psychologique. À moyen ou long terme, l'horizon d'un potentiel effondrement généralisé des économies et des sociétés modernes constitue pour certains « autonomistes » (partisans de l'autosuffisance) un référentiel d’action plausible, incitant à des initiatives préventives et à un changement majeur dans l'interaction avec leur communauté locale \citep{Sevigne-BOOK2015}. 

Les recherches en sociologie se désintéressent des connaissances techniques mobilisées dans l'autosuffisance et de leur production/circulation, pour insister sur les formes d'organisation collective, les engagements politiques, les motivations, les profils sociaux, les parcours de vie, les interactions entre néoruraux et locaux, par exemple. Or, comme une littérature très abondante en STS (Études des Sciences et de la Technologie) le montre, les techniques (agricoles, de construction, etc.) et les formes d'organisation sociale ne peuvent pas être séparées pour être comprises dans leurs évolutions respectives, car elles s'influencent réciproquement et évoluent ensemble.

Dans le cadre du projet CNRS STAY ({\it Savoirs Techniques pour l'Auto-suffisance, sur YouTube}), nous nous intéressons à la caractérisation de la dimension technique, qui viendra rejoindre l'analyse de la dimension sociale. En raison de la tradition de recherche sociologique et de l'approche méthodologique ethnographique le plus souvent choisie, la présence des « autonomistes » sur le Web, et sur YouTube en particulier, n'est pas à ce jour traitée dans la littérature. L'alliance entre analyse sociologique et traitement automatisé des contenus en ligne a un grand potentiel de produire des résultats qui changent complètement de perspective sur l'ampleur et la dynamique du mouvement vers l’autosuffisance en France.  

Nous nous intéressons à l'analyse des données vidéos de YouTube transcrites et des commentaires associés traitant des pratiques en agriculture alternative décrites par des « autonomistes ». La plateforme Agro-STAY a ainsi été proposée pour collecter, traiter, classer et visualiser les données liées aux vidéos et faciliter le travail des sociologues. Des approches fondées sur des modèles d'apprentissage automatique (approches supervisées) et les modèles de langue ont été mobilisées pour une analyse approfondie des données textuelles issues de YouTube. 

La section \ref{etat_art} de cet article résume l'état de l'art des méthodes de collecte et analyse des données YouTube. Les données traitées dans le projet STAY sont décrites en section \ref{sec:ref}. Ces données sont collectées et traitées à travers la plateforme Agro-STAY présentée en section \ref{logiciel}. Des expérimentations effectuées et perspectives sont proposées en sections \ref{expes} et \ref{conclusion}. Il est important de mentionner que cette plateforme est encore en cours de développement, nous évoquons ici sa première version.


\section{Etat de l'art} \label{etat_art}


YouTube, créé en 2005, transgresse les frontières et fournit des contenus vidéos à une large population mondiale. En effet, YouTube est devenu rapidement très populaire auprès des utilisateurs du web: il héberge des vidéos les plus populaires au monde. En 2021, en France, 70~\% des utilisateurs, entre 15 et 24 ans, visitaient le site quotidiennement, y passant en moyenne plus d'une heure par jour, tandis que ceux âgés entre 25 et 49 ans y consacraient 30 minutes par jour \citep{Cervety-2021}. De plus, 70~\% de Français affirment que YouTube a du contenu qu’ils ne trouvent pas ailleurs.
En 2023, YouTube se positionne comme la troisième plateforme la plus visitée sur Internet après Google et Facebook \citep{themedialeader-2023}.

YouTube propose des vidéos couvrant une énorme variété de thématiques, de points d'intérêt et de points de vue.
Ces vidéos, de même que les commentaires qu'elles génèrent, attirent l'attention des chercheurs de différents domaines et disciplines. Mentionnons par exemple un travail sur l'analyse des vidéos dédiées à la santé animale \citep{Bruhl-PHD2023}, une étude des commentaires postés sur différentes vidéos YouTube, comme l'analyse thématique ou des sentiments \citep{Pirnau-ECAI2024}, ou la génération automatique des labels pour la posture d'humains sur les vidéos de YouTube \citep{Dill-EMBC2023}.


\section{Collecte des données dans Agro-STAY} \label{logiciel}

\subsection{Collecte et transcription de vidéos YouTube}

\begin{figure}
\centering
\begin{subfigure}{0.45\textwidth}
    \includegraphics[width=\linewidth]{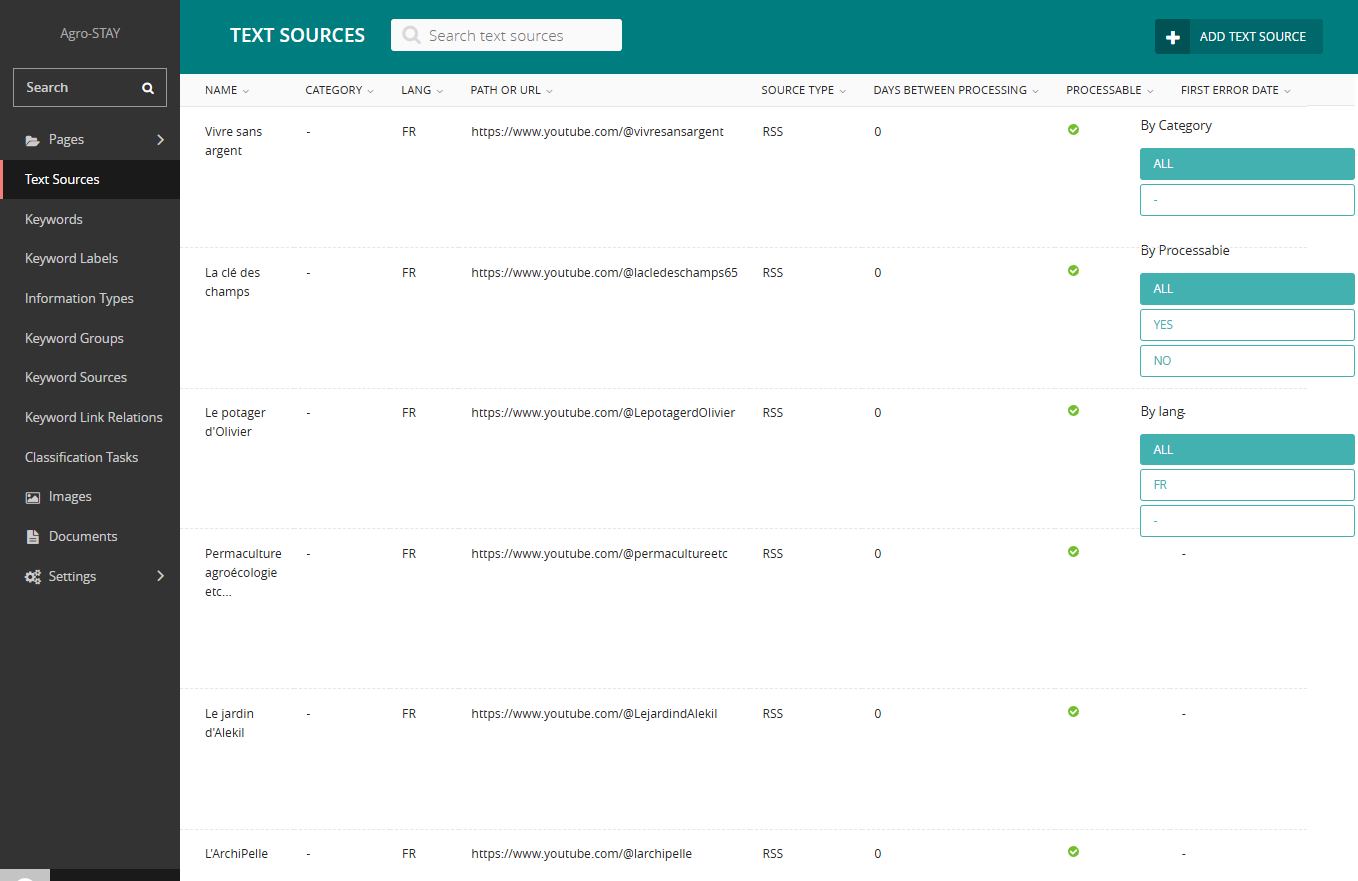}
    \caption{Interface d'administration des chaînes YouTube.}
    \label{fig:liste_sources}
\end{subfigure}
\hfill
\begin{subfigure}{0.45\textwidth}
    \includegraphics[width=\linewidth]{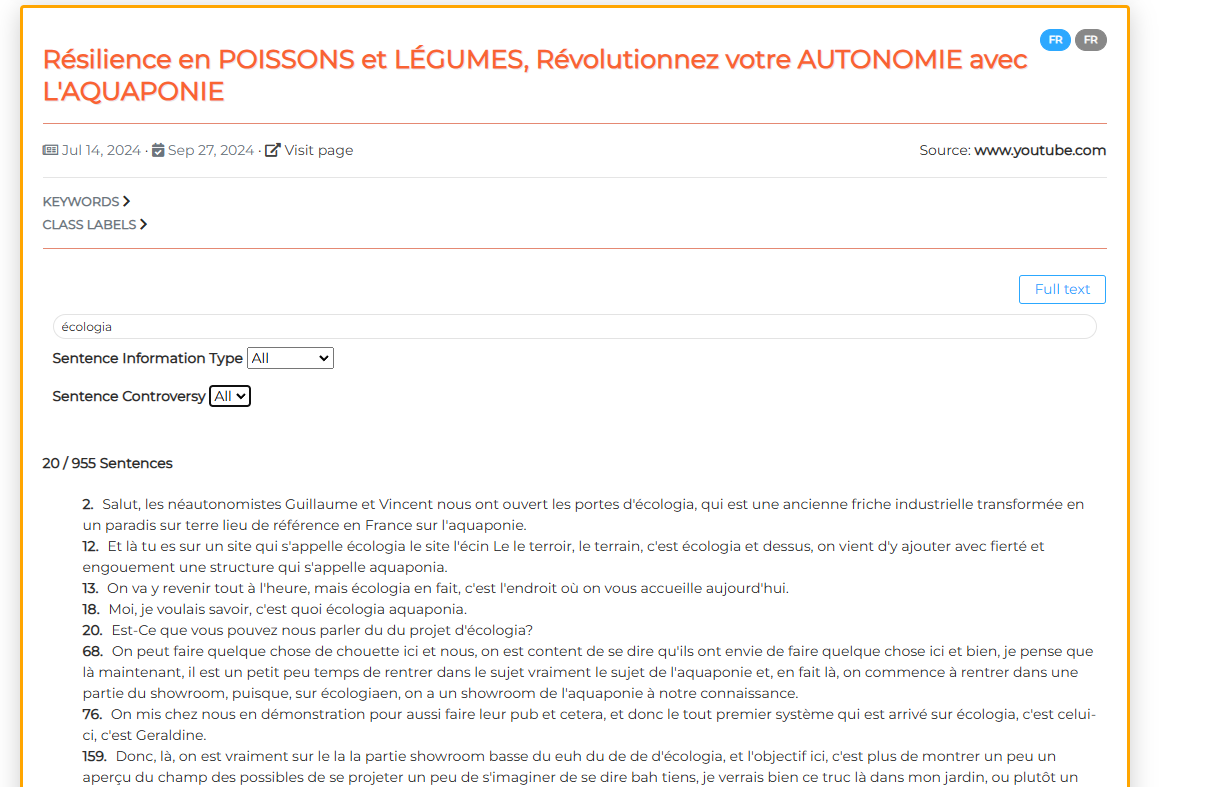}
    \caption{Interface de recherche parmi les phrases d'une transcription.}
    \label{fig:texte_phrases}
\end{subfigure}
\caption{Captures d'écran d'Agro-STAY.}
\end{figure}

Six chaînes YouTube sont présélectionnées par les experts. Toutes ces chaînes promeuvent explicitement l'autonomie alimentaire.
Cet ensemble de chaînes est donné en figure \ref{fig:liste_sources}. Celle-ci présente également l'interface d'administration des sources d'Agro-STAY.

Dans un premier temps, Agro-STAY, pour cet ensemble de chaînes YouTube donné, collecte les transcriptions des dernières vidéos publiées. Cette collecte est réalisée en combinant l’API officielle fournie par YouTube pour la récupération des listes de vidéos d'une chaîne et des méta-données associées, ainsi qu'une librairie tierce\footnote{\texttt{https://pypi.org/project/youtube-transcript-api/}} pour la récupération de la transcription textuelle de la vidéo (ce qui n'est pas possible avec l'API officielle). Agro-STAY intègre également les mécanismes de mise à jour automatique des données.
Les transcriptions des vidéos de YouTube ont plusieurs limites : elles contiennent de nombreuses erreurs de transcriptions et ne comportent aucune mise en forme (absence de ponctuation, majuscules, etc.). Ces particularités entraînent des défis pour les techniques et modèles de TAL.
Nous utilisons la librairie python Punctuator \footnote{\texttt{https://pypi.org/project/punctuator/}} pour la restitution de la ponctuation et des majuscules.
D'autres traitements appliqués sont, en particulier : la segmentation en phrases, une classification automatique, une indexation d'entités nommées et de mots-clés prédéfinis (lieux, organisations...). Cette fonction exploite la librairie spaCy.
Les textes sont consultables via l'interface d'Agro-STAY, avec leurs méta-données (dates de publication et de collecte, chaîne YouTube d'origine, URL de la vidéo). La transcription peut être visualisée sous sa forme textuelle brute ou sous la forme d'une liste de phrases (figure~\ref{fig:texte_phrases}).
Actuellement, 1~423 vidéos sont collectées et traitées. 

\subsection{Collecte et traitement des commentaires}

\begin{figure}
\centering
\begin{subfigure}{0.45\textwidth}
    \includegraphics[width=\linewidth]{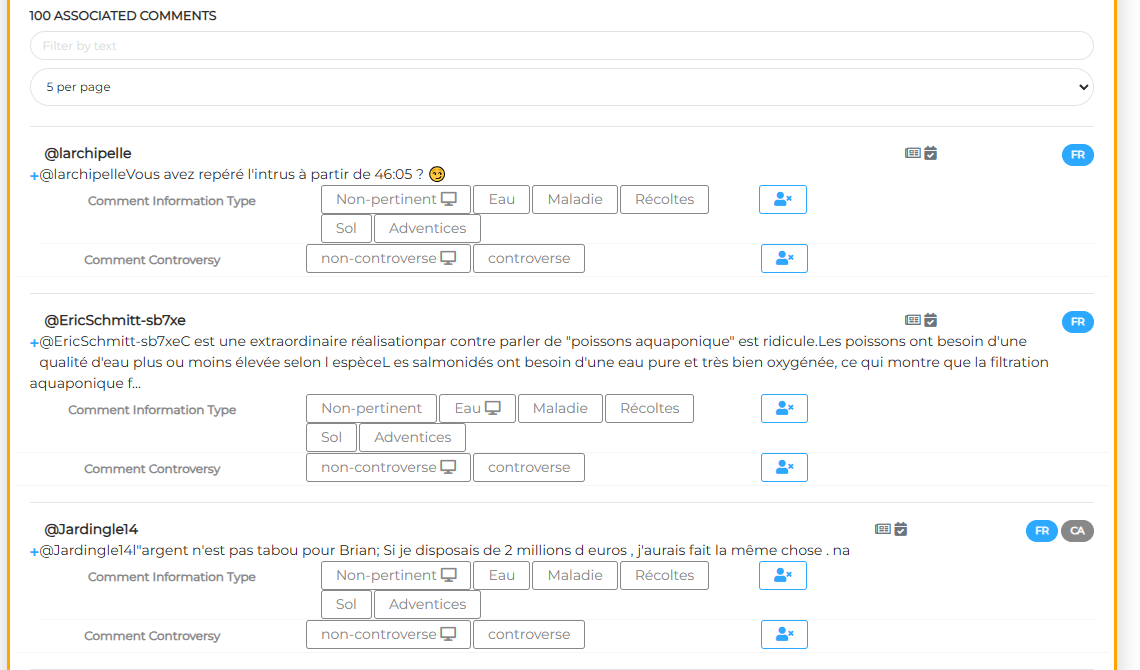}
    \caption{Consultation des commentaires associés à une vidéo.}
    \label{fig:commentaires}
\end{subfigure}
\hfill
\begin{subfigure}{0.45\textwidth}
    \includegraphics[width=\linewidth]{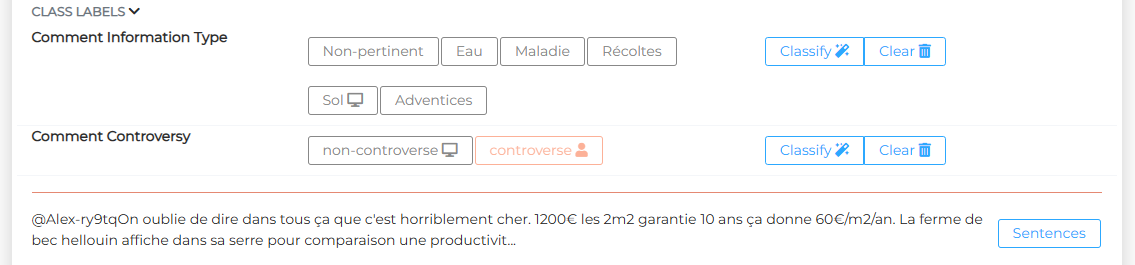}
    \caption{Annotation manuelle d'un article dans la classe \textit{Controverse}.}
    \label{fig:texte_phrases}
\end{subfigure}
\caption{Captures d'écran d'Agro-STAY.}
\end{figure}

Les commentaires associés à chaque vidéo sont également collectés avec l'API officielle de YouTube. La particularité des commentaires est qu'ils sont très nombreux et qu'ils comportent des smileys, fautes d'orthographe ou de grammaire.
Actuellement, nous avons collecté près de 45~000 commentaires. 
Les commentaires sont traités et indexés selon les mêmes étapes que les transcriptions de vidéos (segmentation en phrases, indexation, classification, etc.), ce qui offre les mêmes capacités de recherche parmi les commentaires.
L'interface d'Agro-STAY permet de consulter les commentaires associés à une vidéo (figure~\ref{fig:commentaires}).

\subsection{Préparation des données de référence}
\label{sec:ref}

Une première annotation a été réalisée par deux sociologues experts en techniques agricoles de l’autonomie. Un expert a annoté un échantillon de 64 vidéos (pour un total de 20 heures) et 100 commentaires associés à 12 vidéos, tandis que le second a annoté un sous-échantillon de 5 vidéos (pour un total de 7 heures et 43 minutes) et 1400 commentaires associés à 12 vidéos. Les catégories possibles sont : (1) maladies et ravageurs, (2) gestion de l’eau (paillage, irrigation, ...), (3) adéquation entre qualités du sol et besoins des plantes cultivées, (4) gestion des adventices.
Après ce premier tour d'annotations, avec la création d'un guide et la description des classes, les experts ont discuté les désaccords des annotation et affiné les définitions des classes. 
Le tableau \ref{tab:classes} présente les définitions des classes et fournit des exemples.

\begin{table}[h!]{\footnotesize
\centering
\begin{tabular}{|p{2.8cm}|p{5cm}|p{6cm}|}
\hline
\textbf{Classe} & \textbf{Définition} & \textbf{Exemples (extrait de vidéo et commentaire)} \\
\hline
Gestion des maladies et des ravageurs (nom court : \textit{Maladies/Ravageurs})&
Cette classe inclut les actions mises en place pour gérer les maladies et ravageurs des plantes cultivées (arbres compris), comme les limaces, les pucerons, les maladies fongiques, etc.. Peut inclure, par exemple, des actions telles que le traitement avec bouillie bordelaise, les traitements par des purins, le lâcher de poules ou de canards indiens, le choix des porte-greffe ou des variétés résistantes, l’organisation spatiale des cultures pour éviter la concentration d’une même espèce, la plantation d’espèces végétales réputées protectrices. &
\textbf{Vidéo} : \textless il y aura tous les poireaux qui vont attirer tous les ravageurs qui raffolent des poireaux. Bah on va mettre un peu de poireaux là, un peu de poireaux là, un peu de poireaux là. Et puis en remettre un chou et puis là alors moi je fais quand même des vrais rangs, mais j'essaie de qui a un petit peu de variété. J'essaie de mélanger légumes entre eux, de tenir compte aussi de leur forme et de leur taille. Là, on a un exemple qui est pas mal justement, je suis juste à côté donc là c'est 2 rangs vraiment très bête \textgreater
~\textbf{Commentaire} : \textless C'est super tout ça mais qd les rats taupiers sont là, cette énergie déjà testée se termine en fiaco, en échec total, en épluchures. C'est du vécu HORRIBLE \textgreater\\
\hline
Gestion de l’eau (paillage, irrigation, strates végétales, eau du réseau, de pluie ou de source  (nom court : \textit{Eau})&
Cette classe inclut les actions mises en place pour gérer l’eau dans la production agricole – qu’il s’agisse d’accès à l’eau (puits, source, réseau d’eau potable, ...), de manque d’eau (sécheresse..), d’excès d’eau (sols hydromorphes..), les économies d’eau pour éviter le gaspillage et réduire les coûts. Ces actions peuvent être, par exemple, la mise en place de paillage pour éviter l’évaporation en été, l’irrigation par goutte-à-goutte, etc. &
\textbf{Vidéo} : \textless  J'ai pas d'eau ici hein quasiment pas. J'ai 2 cuves ridicules qui me permettent, qui me permettaient d'arroser la serre. Et puis un peu le démarrage de plantation donc évidemment ça a pas été arrosé comme la terre est pas très riche \textgreater
~\textbf{Commentaire} : \textless Je pense que tu n'as pas mentionné l'arrosage sous serre qui peut etre problématique si tu veux planter en pleine terre ? A+ \textgreater\\
\hline
Gestion de l’adéquation entre qualités du sol et besoins des plantes cultivées  (nom court : \textit{Sol})&
Cette classe inclut des actions mises en place pour rendre le sol le plus en adéquation possible avec les besoins des plantes cultivées, en termes de structure, composition et fertilité. Quelques exemples de telles actions sont l’apport d’engrais organique ou pas (fumiers divers, engrais de synthèse ; apport de compost) ; l’apport de biomasse en paillage (paille, foin, broyat, brf, feuilles mortes, branchages coupés pour cet objectif) ; favoriser le vivant dans le sol ; pratiquer la perturbation des végétaux « sacrificiels » pour stimuler les végétaux autour d’eux. & 
\textbf{Vidéo} : \textless  Sauf que moi … c’est trop long. Je veux du tropical quoi, tout de suite je la jungle et c'est basé sur le principe de… on va couper au centre toute la biomasse et la rejeter sur les bords pour faire des boudins comme les boudins de paille qu'on a qui vont protéger l'enracinement et on va utiliser la matière aérienne ligneuse pour l'apport de carbone, principalement pour l'apport de carbone qu'on va ramener au sol. [...] Le bananier ou les arbres d'apport de biomasse ligneuse sont pléthoriques au Brésil, ça pousse à une vitesse folle, mais pas chez nous \textgreater
~\textbf{Commentaire} : \textless La grelinette, outil magique pour décompacter la terre et la houe, outil de nos grands-parents très efficace pour affiner le travail. Bon courage pour la suite.
 \textgreater\\
\hline
Gestion des adventices  (nom court : \textit{Adventices})& 
Cette classe inclut des actions mises en place pour gérer les végétaux indésirables, aussi nommées « mauvaises herbes » ou adventices. Des exemples incluent l’utilisation de bâches en tissu renforcé, la mise en place de couvert végétaux, la densification des cultures, le désherbage manuel, mécanique ou chimique. &
\textbf{Vidéo} : \textless  Depuis deux ans, je ne désherbe même plus. C'est avant, je continuais à désherber. Même si je désherbais, je laissais sur place et je paillais. Aujourd'hui, je passe comme ça. Je regarde, il y a des pics parfois. Je passe juste et j'écrase au final. J'enlève un peu ce qui gêne la lumière. Et rien que le fait de venir aplatir au final, celui-là qui est là, hop, il vient aplatir et avant que ça se relève, ça va mettre déjà un petit mois. [...] \textgreater
~\textbf{Commentaire} : \textless Si le liseron apparait, voir du côté du sol pour déterminer la carence ou le surplus qui déclenche la dormance.
 \textgreater\\ 
\hline
Récolte  (nom court : \textit{Récolte})& 
Cette classe est exclusive aux commentaires, elle inclut ce qui concerne les récoltes et n'est pas couvert par d'autres classes. &
\textbf{Commentaire} : \textless  Dans le Maine-et-Loire, superbe récolte cette année. Un pied de chayotte et une centaine de fruits. \textgreater \\
\hline
\end{tabular}
}
\caption{Description des classes et exemples associés.}
\label{tab:classes}
\end{table}





\section{Classification} \label{expes}




Deux tâches de classification sont définies :
\begin{itemize}
    \item \textit{Classification des types d'information} selon les six classes décrites dans le tableau~\ref{tab:classes} : \textit{Maladies/Ravageurs}, \textit{Eau}, \textit{Sol}, \textit{Adventices}, \textit{Récolte}\footnote{Cette classe n'est présente que dans les commentaires. Elle couvre tout ce qui concerne la description de récoltes.} et \textit{Non-pertinent} (le reste).
    \item \textit{Classification de la controverse} pour détecter les commentaires controversés, selon une classification binaire (\textit{controverse} et \textit{non-controverse}).
\end{itemize}

Les modèles de classification entraînés visent à être appliqués à la fois sur : (1) les commentaires associés aux vidéos, pour sélectionner les commentaires d'intérêt; (2) les phrases des transcriptions, afin d'en cerner les fragments pertinents; (3) les transcriptions complètes.

\subsection{Données d'apprentissage}

Pour l'entraînement des modèles, nous utilisons 1~400 commentaires annotés selon les thématiques (section \ref{sec:ref}) et la présence de controverses. Ces données sont plus faciles à exploiter que les données de transcriptions de vidéos, notamment en raison de leur brièveté.

\begin{figure}
\centering
\begin{subfigure}{0.45\textwidth}
    \includegraphics[width=\linewidth]{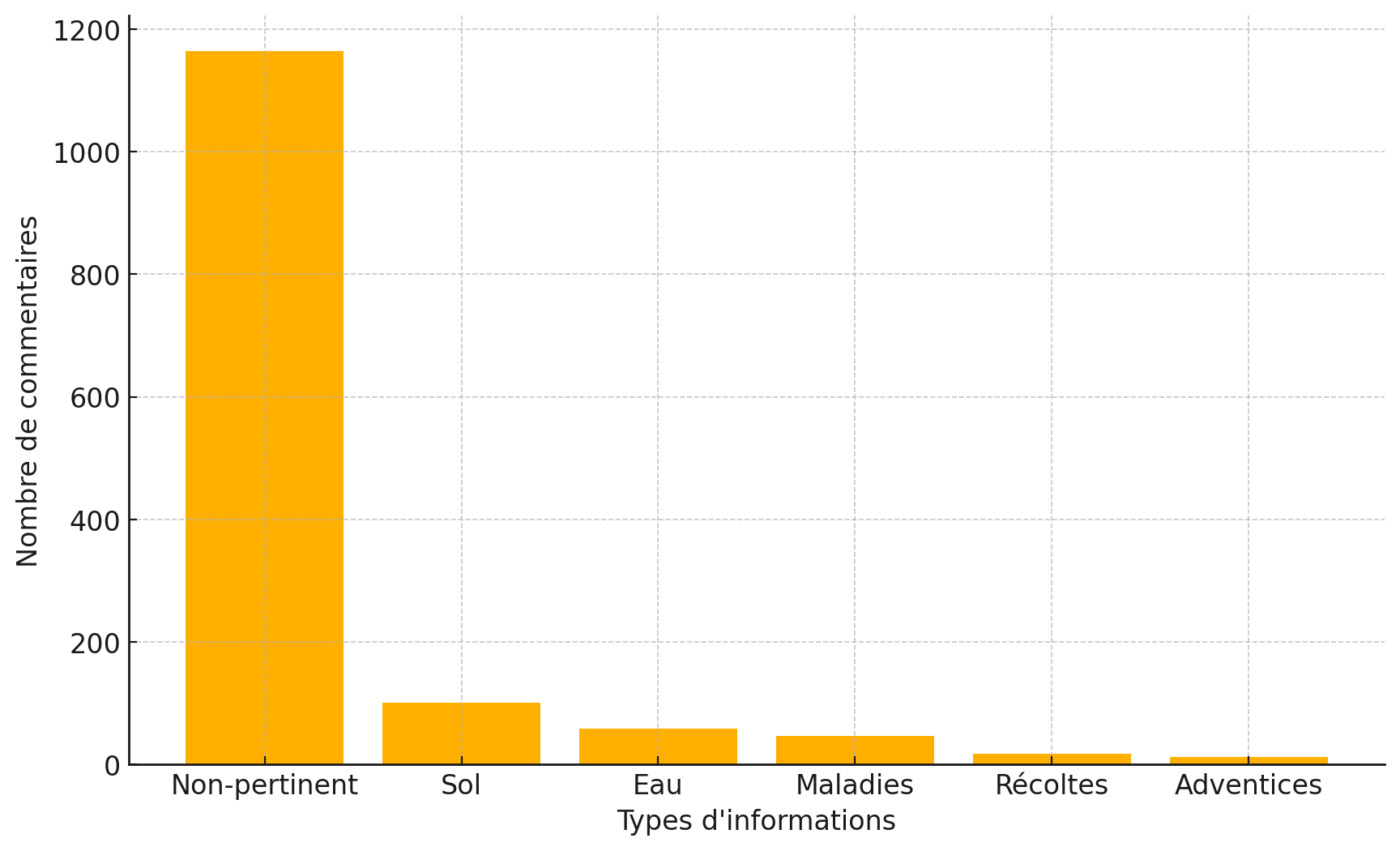}
    \caption{Annotations \textit{Type d'information}.}
    \label{fig:comment_information_type_distribution}
\end{subfigure}
\hfill
\begin{subfigure}{0.45\textwidth}
    \includegraphics[width=\linewidth]{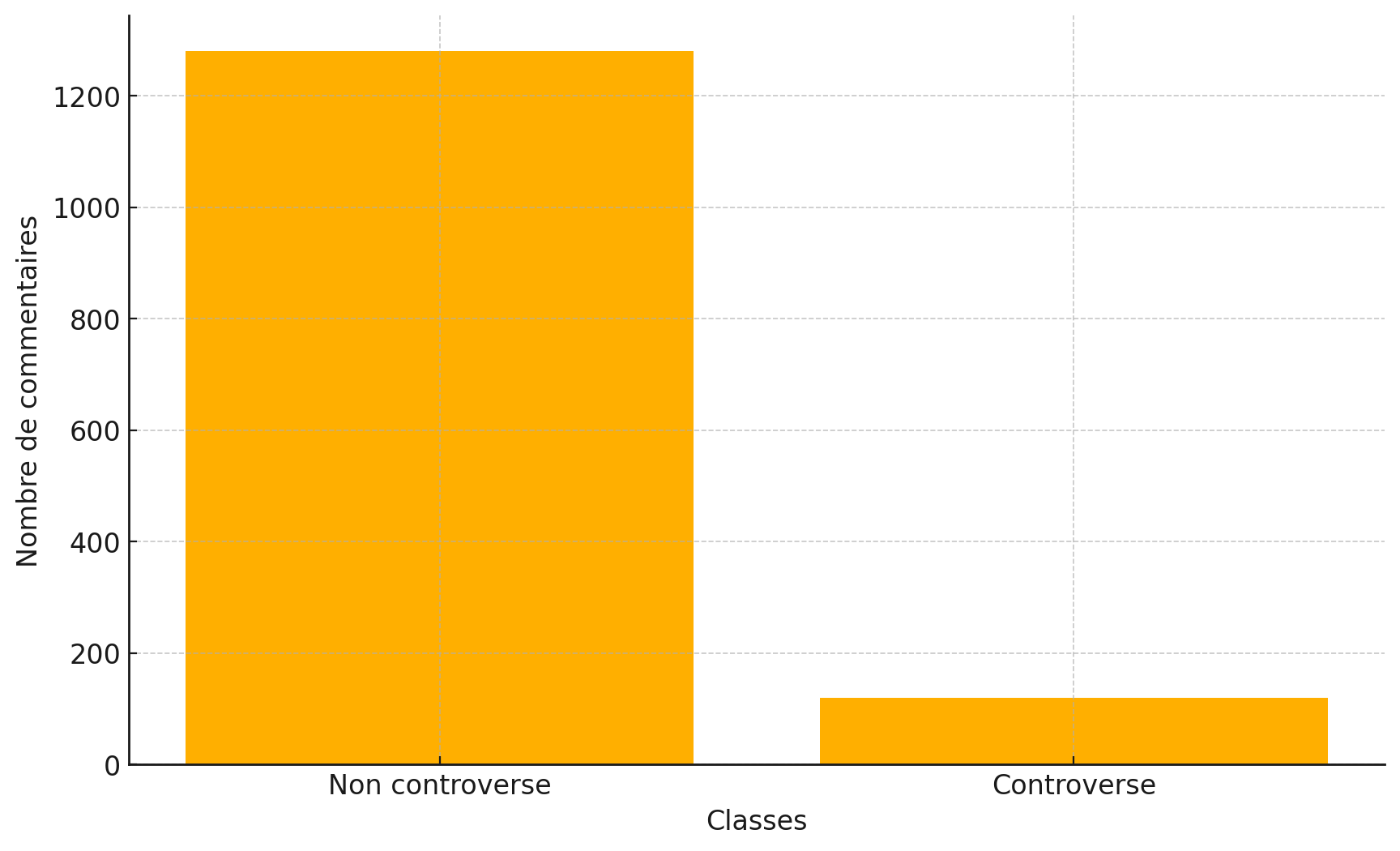}
    \caption{Annotations \textit{Controverse}.}
    \label{fig:comment_controversy_distribution}
\end{subfigure}
\caption{Distribution des classes dans les données annotées.}
\end{figure}

Le travail sur les commentaires présente plusieurs défis : il s'agit généralement de textes courts et peu formatés, comportant parfois des smileys, des abréviations, ainsi que des fautes d'orthographe ou de grammaire. 
De plus, les données sont très déséquilibrées (environ $90\%$ de commentaires \textit{non-pertinents} ou \textit{non-controverse}), comme le démontrent les figures \ref{fig:comment_controversy_distribution} et \ref{fig:comment_information_type_distribution}.

\subsection{Apprentissage des modèles de classification}

Nous utilisons le modèle de langue pré-entraîné \textit{CamemBERT} \citep{martin-etal-2020-camembert}. Ce modèle est basé sur l'architecture RoBERTa \citep{2019roberta} et est optimisé pour le français grâce au corpus multilingue OSCAR \citep{2022arXiv220106642A}.
Nous utilisons ce modèle après un ré-apprentissage (\textit{fine-tuning}) sur les commentaires annotés. Cependant, aucun prétraitement approfondi ni optimisation des hyperparamètres n'ont été réalisés. Les paramètres utilisés pour l'apprentissage sont les valeurs par défaut proposées dans la documentation\footnote{\texttt{https://almanach.inria.fr/software\_and\_resources/CamemBERT-fr.html}}.
Les données sont segmentées en deux sous-ensembles : $80\%$ pour l'apprentissage, $20\%$ l'évaluation.

\subsection{Résultats}

\begin{table}
\centering
\begin{tabular}{|c|c|c|c|c|}
\hline
\multirow{2}{*}{\textbf{Jeu de données}} & \multicolumn{4}{c|}{\textbf{Scores}} \\
\cline{2-5}
 & \textbf{Précision} & \textbf{Rappel} & \textbf{F1} & \textbf{Exactitude} \\
 & \textit{{\footnotesize macro (pondéré)}} & \textit{{\footnotesize macro (pondéré)}} & \textit{{\footnotesize macro (pondéré)}} &  \textit{{\footnotesize(accuracy)}}\\
\hline
Controverse & 70.54 (92.2) & 73.81 (91.52) & 72.01 (91.83) & 91.52 \\
\hline
Type d'information & 41.23 (84.14) & 40.52 (85.27) & 40.57 (84.62) & 85.27 \\
\hline
\end{tabular}
\caption{Performances selon les classes (valeurs macro et pondérées).}
\label{tab:classification_results}
\end{table}

La performance des modèles est évaluée avec la précision (macro et pondérée), le rappel (macro et pondéré), le F1-score (macro et pondéré), et l’exactitude globale. Les résultats obtenus sont présentés dans le tableau~\ref{tab:classification_results}.
Les résultats montrent une différence significative entre les scores macro et pondérés. Cette disparité met en évidence l'impact du déséquilibre des classes dans les jeux de données : les classifieurs tendent à privilégier les classes sur-représentées, ce qui entraîne des scores pondérés plus élevés que les scores macro.

La classification \textit{Controverse}, qui est binaire (\textit{controverse} et \textit{non-controverse}), a des résultats relativement élevés avec un F1-macro de 72.01. Toutefois, cette tâche est également affectée par le déséquilibre des données. Cela peut expliquer les difficultés rencontrées pour maintenir des performances équilibrées entre les deux catégories.

La classification \textit{Type d'information}, qui comporte six classes, s’avère plus complexe en raison du plus grand nombre de catégories. Malgré un F1-macro de 40.57, les performances dépassent largement celles d’un classifieur aléatoire, ce qui démontre une capacité du modèle à distinguer les classes dans les données, bien qu'une grande marge d'amélioration existe.


Ces résultats, bien que perfectibles, sont encourageants et indiquent des pistes d'amélioration des performances des modèles, comme : (1) rééquilibrage des données pour réduire l'impact des classes sur-représentées, (2) meilleurs prétraitements, notamment pour nettoyer et normaliser les textes courts, informels ou mal orthographiés, (3) optimisation systématique des paramètres d’apprentissage pour maximiser l’efficacité du modèle.

\section{Conclusion et perspectives} \label{conclusion}
Le projet STAY met en lumière l'importance d'une approche pluridisciplinaire pour étudier le mouvement croissant vers l'autosuffisance en France. En articulant les dimensions sociologiques et techniques, il montre comment des savoirs techniques, largement diffusés via des plateformes numériques comme YouTube, se combinent aux motivations et aux organisations sociales des « autonomistes ». 
La plateforme Agro-STAY, en exploitant les avancées récentes en traitement automatique des langues et en apprentissage automatique, permet d'accompagner les experts vers une analyse plus systématique des pratiques agricoles alternatives et des interactions sociales autour de celles-ci. 
Il convient, dans les futurs travaux liés au projet STAY, d'approfondir le traitement de données déséquilibrées en étudiant notamment les approches d'augmentation de données.
Aussi, les résultats expérimentaux présentés dans cet article concernent principalement les résultats obtenus sur les commentaires, les futurs travaux étudieront également la classification des textes (phrases) issues des transcriptions YouTube.
Les méthodes sont génériques et pourront être étendues à d'autres thématiques de façon aisée.

\section*{Remerciements} 
Ce projet a obtenu le soutien financier du CNRS à travers les programmes interdisciplinaires de la MITI. Nous  l’Agence nationale française de la recherche dans le cadre du programme Investissements d’avenir \#DigitAg, référencé ANR-16-CONV-0004.

\bibliographystyle{rnti}
\bibliography{biblio_exemple}

\end{document}